\documentclass{sig-alternate}

\usepackage{url}

\usepackage{algorithm}
\usepackage{algpseudocode}
\usepackage{topcapt}
\usepackage{booktabs}
\usepackage{color}

\usepackage{etoolbox}
\makeatletter
\patchcmd{\maketitle}{\@copyrightspace}{}{}{}
\makeatother

\usepackage{textcomp}

\newcommand{\mpara}[1]{\medskip\noindent{\bf{#1}}}

\usepackage{multicol}

\usepackage{mathptmx}

\usepackage{amsmath}

\usepackage{amssymb}

\usepackage{graphicx}

\newif\ifnotes
\notesfalse
\ifnotes
\newcommand{\authnote}[2]{{ $\ll$\textsf{\footnotesize #1: #2}$\gg$}}
\else
\newcommand{\authnote}[2]{}
\fi

\begin{document}
\thispagestyle{empty}


\title{On Spectral Analysis of the Internet Delay Space and Detecting Anomalous Routing Paths}
\author{
\centerline{{Gonca G\"ursun}}\\
\centerline{\sc{Department of Computer Science, Ozyegin University}}
}
\maketitle

\begin{abstract}
Latency is one of the most critical performance metrics for a wide range of applications. Therefore, it is important to 
understand the underlying mechanisms that give rise to the observed latency values and diagnose the ones that are 
unexpectedly high. In this paper, we study the Internet delay space via robust principal component analysis (RPCA). 
Using RPCA, we show that the delay space, i.e. the matrix of measured round trip times between end hosts, 
can be decomposed into two components - the expected latency between end hosts with respect to the current state 
of the Internet and the inflation on the paths between the end hosts. Using this decomposition, first we study the
well-known low-dimensionality phenomena of the delay space and ask what properties of the end hosts define the dimensions. Second, using the decomposition, we develop a filtering method to detect the paths which experience 
unexpected latencies and identify routing anomalies. We show that our filter successfully identifies an 
anomalous route even when its observed latency is not obviously high in magnitude. 

\end{abstract}

\section{Introduction}    \label{sec:intro}   Latency is one of the most important performance metrics.
The quality of a wide range of applications, such as server
selection in CDNs, video streaming, voice over IP, as well as any
time-critical application, require low latency on the Internet
paths. Therefore there has been great interest to understand the 
root causes of high RTT values \cite{Krishnan09, Pucha07, Zarifis14}. 

Beside the physical distance between two end hosts, one key factor
that drives the latency is routing. Both intradomain and 
interdomain routing decisions of ASes on the paths impact the latencies 
\cite{Krishnan09, Spring03, Tangmunarunkit01}. In fact, the impact of 
routing on latency is two-fold. First, phenomenas behind routing decisions, 
together with the physical distance between end hosts, generate patterns in 
the latency data and result in a low-dimensional delay space \cite{Abrahao08, Tang03}. 
In other words, a matrix of RTT values between end hosts is effectively low-rank 
\footnote{A full-rank matrix is effectively low-rank if the matrix can be well-approximated by its first few principal components.}. 
Second, suboptimal routing choices and misconfigurations can increase the latencies on the paths and possibly
cause discrepancies in the matrix structure. Leveraging these observations, in this paper, 
we propose a spectral decomposition of latency matrices to distinguish the regular structure 
of the latency matrix from the discrepancies. Our goal is to write a given latency matrix as a linear 
combination of two matrices: a low-rank \emph{expected} latency matrix that reveals structure in the delay space
and an \emph{inflation} matrix that reveals the noise and the inflation in RTTs that does not fit into the low-rank structure.   
We decompose latency matrices via a recently developed technique, Robust PCA \cite{Candes09}, as described in Section~\ref{sec:rpca}.

Using this decomposition, we aim to answer the following questions. First, we ask what properties of a given latency matrix 
contribute to its overall dimensionality. To answer this question, we study the rank values of  a bunch of \emph{expected} latency matrices
and investigate whether there is a correlation between their rank values and some features of the end hosts. We find that 
the number of unique AS-geolocation of the end hosts on the rows/columns of the matrix determines its rank. 
Second, we ask whether we can detect anomalous routing paths via our decomposition. We mark any path to be an 
\emph{anomaly candidate} if a significant portion of its RTT is estimated as inflation. In other words, 
for each path, we compute the ratio of the inflation to the expected latency. If this ratio is higher 
than a threshold, we investigate the path as a possible anomaly. 

Notice that our definition of inflation is different than the previous work
in which a path is called inflated if the measured RTT is significantly greater than the 
lower bound RTT computed based on the physical distance. Often times the routing paths
are not the physical shortest paths and actual speeds of packets are much slower than 
the theoretical speed of mediums. Therefore, comparing the observed RTTs with the lower bounds do not 
pinpoint anomalous routes unless the RTT is obviously much larger than the lower bound. Unlike the 
previous work, our approach does not set lower bounds. We compute the expected latency on a path 
with respect to the current delay state of the Internet via our decomposition. Then we decide whether a path is 
an anomaly candidate or not by comparing its expected RTT component with its inflation component.

We summarize our contributions in this paper as follows: a) we show how to decompose a latency matrix 
via a recent technique, RPCA, into a low-rank and an inflation component, b) we investigate 
the features of end hosts that result in the low-rank property and we find that
both geolocation and the AS of the end hosts define the dimensions of the delay space, c) we propose
a method to diagnose the inflated paths by the inflation-to-expected-latency ratio filters, d) we show
that our filter successfully pinpoints the routing anomalies even when the 
RTT on the paths are not obviously large, e) we show that our filter successfully 
pinpoints the routing anomalies even when all measured paths between the end hosts from two regions
are inflated, f) we show how to apply our filter in case RTT measurements
between some end hosts are missing.

The rest of the paper is organized as follows. We
introduce the spectral analysis tool, RPCA, in Section~\ref{sec:rpca}  and describe
our dataset in Section~\ref{sec:dataset}. In Section~\ref{sec:rank} we study why the delay space is 
low-dimensional. In Section~\ref{sec:cases} we present the anomalous routes that we detect in our dataset.  
We present the related work in Section~\ref{sec:rel-work} and conclude in Section~\ref{sec:conc}.

\vspace{-0.1in}

\section{Decomposing Latency}   \label{sec:rpca}  Latency matrices are shown to be effectively low-rank in the previous work. This property of latency matrices are used in
various applications such as embedding the delay space into low-dimensional coordinate spaces and estimating RTTs 
between hosts without direct measurements \cite{Abrahao08, Dabek04, Liao12, Madhyastha06, Mao04, Tang03, Wong05, Zhang06}. 
Our goal in this paper is also to leverage the low-rank property in order to distinguish the end hosts 
that experience expected latencies with respect to the current state of the delay space from the end hosts that experience
unexpected latencies. 
The first step to do that is to find a low-rank approximation of a given latency matrix. 

Principal Component Analysis (PCA) is the most popular tool to find a low-rank approximation for a given matrix 
and widely applied on latency matrices. Despite its popularity, PCA has a few drawbacks - it is highly sensitive to arbitrarily large or grossly corrupted observations 
and missing measurements \cite{Ringberg07}. Such cases are quite common in RTT measurements, e.g. due to incomplete
traces by non-responsive end hosts, system limitations and failures result in missing and corrupted observations, anomalies
cause arbitrarily large RTT values. PCA is \emph{not robust} to such cases, i.e. it fails to 
yield the true underlying structure of the data. A recent technique called Robust PCA (RPCA) 
addresses these drawbacks \cite{Candes09} and suits well our decomposition goals. 

\mpara{RPCA.} Let $X$ be a data matrix. In our context, the rows of $X$ are the sources, the columns are the destinations, and 
an element stores the RTT value observed between the corresponding source and the destination. RPCA aims to find 
a decomposition of $X$ such that $X$ = $L$ + $S$. In this decomposition, $L$ is a low-rank matrix generated by 
the underlying mechanism of the observed data and $S$ is a sparse noise matrix that does not fit into the low-rank
property. In our context, $L$ stores the expected latency values between end hosts and $S$ stores the unexpected 
inflation (one can also call the values of $S$ as noise). 

RPCA decomposes a given $X$ into its $L$ and $S$ under the following conditions:
a) The rank or the column and row spaces of $L$ are unknown,
b) The number of non-empty entries of $S$ is unknown,
c) The locations of non-empty entries of $S$ are unknown,
d) The entries of $L$ and $S$ can be arbitrarily large,
e) The non-empty entries of $S$ are randomly distributed.
In order to find $L$ and $S$ under these conditions, RPCA solves the following optimization problem:

\begin{equation}
	\begin{aligned}
	& \text{minimize}
	& & ||L||_{*} +  ||S||_{1}\\
	& \text{subject to}
	& & L + S = X 
	\end{aligned}
\label{eq:rpca}
\end{equation}

where $||L||_{*}$ is the nuclear norm of matrix $L$ which is defined as the sum of its 
singular values and $||S||_{1}$ is the $l_1$ norm of $S$. \cite{Candes09} shows that solving
this optimization problem can exactly recover $L$ and $S$ in polynomial time. Note that writing 
the optimization over $l_1$ norm and the nuclear norm is one of the keys to deal with the arbitrarily large, corrupted,  
and missing data. On the other hand, traditional PCA optimizes over $l_2$ norm which makes it
sensitive to the large, corrupted, and missing data points. 
In our analysis, we use the implementation of RPCA provided by the authors of \cite{Candes09}
and refer the reader to their paper for further detail on the tehcnique.

Notice that RPCA is not just a variant of PCA. Instead, the difference between two methods is substantial. 
While both techniques aim at identifying the core low-rank component $L$ from a given measurement matrix $X$,
they do it under completely different assumptions about the additive perturbation. PCA assumes that a gaussian
and non-sparse perturbation $S$ and minimizes $l_2$ norm, while RPCA assumes the perturbation component $S$
to be sparse regardless of its distribution and therefore minimizes $l_1$ norm for $S$ and nuclear norm for $L$.




\section{Datasets}   \label{sec:dataset}  \vspace{-0.05in}
\mpara{RTT data.} We use a collection of RTT values collected from Akamai CDN
\footnote{This work was initiated when the author was visiting Akamai Technologies.}
via traceroute measurements 
on January 24, 2016. The measurements are taken from 47 Akamai server nodes to 5076 
client IPs located in France. The Akamai server nodes are spread 
across 14 unique ASes in three countries, i.e. France, USA, and Japan.

Due to the scale of our measurement setup, there are limitations on the number of times a client IP can be 
tracerouted at a given time period. Such limitations are set by the ISPs in order not to keep client IPs busy. 
Therefore, we tracerouted each client IP from 20 Akamai server nodes on average. That is, we do not have
an RTT measurement for all server node and client IP pairs. 

In practice, there are several issues with using traceroute data. One issue is that ICMP packets might be deprioritized 
or simply dropped at the routers. These result in higher RTTs or incomplete traces. To minimize these effects,
we take three consecutive measurements from a given server node to a client IP. Then, we use the one with the minimum latency.
We also remove all incomplete traces from our dataset.

Another issue is the asymmetric reverse paths and latency inflation due to long reverse paths. In fact, 
we find asymmetry and circuitousness in reverse paths are common. One indication of circuitousness
in the reverse path is the significant increase of RTT on a single hop even though the consecutive routers in the forward path
are geographically close. We discard these cases from our analysis and focus on the anomalies on the forward paths.

\mpara{BGP Announcements.} We use a collection of BGP tables to analyse the inflated paths.
The tables are collected on the same day as traceroute measurements from 297 peer routers in Akamai CDN.  
They consist over 790K BGP paths to over 48K prefixes in Europe. 
Using this dataset, we map our client IPs to their longest matching BGP prefixes. 
5076 client IPs map to 778 prefixes.

\mpara{Geolocation Mapping.} We use the EdgeScape tool of Akamai 
to map each and every IP address 
in our dataset to its geographic location \cite{edgescape}.

\section{Understanding Dimesionality}  \label{sec:rank}  
In this section we ask what makes the latency data low-dimensional. To answer our question we 
decompose latency matrices, s.t. $X$ = $L$ + $S$. Then we study the rank of the low-rank $L$ components. 
Our goal is to find which features of the end hosts correlate with the rank values. 

We generate $X$ matrices in two levels of granularity. 
In the first level, each column of $X$ is an individual destination IP. In the second level,
we aggregate individual IPs into their prefixes and each column of $X$ represents a BGP prefix. 

\mpara{IP-level.}
We generate an $X_{IP}$ matrix for each BGP prefix such that the columns are the individual destination IPs that belong to the prefix 
and the rows are the server nodes. From each $X_{IP}$, we discard the rows which have no measurements to any of the destinations on the columns. 
Then, we decompose each matrix $X_{IP}$ into its $L_{IP}$ and $S_{IP}$ and compute the rank of $L_{IP}$. 
 
\begin{figure}[tbp]
\centerline{
\includegraphics[width=0.29\textwidth]{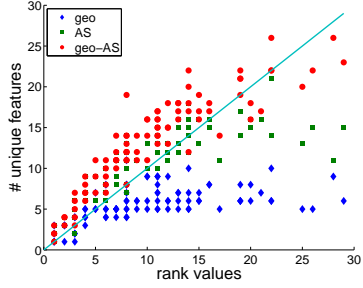}}
  \vspace{-0.04in}
\caption{The number of unique features of the server nodes vs. the rank values of the $L_{IP}$ matrices.} 
\label{fig:iplevel_rankL}
\vspace{-0.15in}
\end{figure}

Our intuition is that since the routing decisions are made at the level of BGP prefixes, each column vector of a given
$X_{IP}$ will be same or very similar to each other. Therefore, the rank of $L_{IP}$ matrices will be correlated with the
features of the server nodes in the rows. In order to show that our intuition holds, we plot the rank values of $L_{IP}$ 
matrices in Figure~\ref{fig:iplevel_rankL}. Note that, in order not to limit the rank of a matrix by its number of rows or columns, 
we only consider matrices that has large number of individual IPs, i.e. the number of columns in $X_{IP}$ vary between 50 and 222. 

The Figure~\ref{fig:iplevel_rankL} plots the rank of $L_{IP}$  matrices vs. three features of the server nodes. We tag each server node
by a) its geolocation, b) the AS that the server node belongs to, c) both the geolocation and the AS of the server node. 
First, Figure~\ref{fig:iplevel_rankL} shows that the rank values are relatively low, i.e. they vary between 1 and 30. 
Second we find that the number of unique geolocations or the number of unique ASes of the server nodes are not enough to explain the low dimensionality. 
Instead, the number of unique geolocation and AS pairs in the rows best correlate with the rank values. Next, we ask whether our findings
still hold when we increase the diversity on the columns by bringing various prefixes from different ASes together.

\mpara{Pfx-level.}
There are various advantages in aggregating individual IPs into their prefixes, e.g. less noise, smaller data size. 
Also, such aggregation is natural since the routing in the Internet is based on BGP prefixes, i.e. prefixes are atomic 
with respect to routing. We analyze the dimensionality of the delay space under this aggregation. 

We map all IPs into their BGP prefixes in our dataset.  Then we generate a matrix $X_{FR}$ s.t. the rows are the server
nodes, each column represents one prefix, and an element is the minimum latency measured from the corresponding
server node to all IPs within the corresponding prefix on the column. We consider only the large prefixes, i.e. 
prefixes that has at least 10 IPs mapped to them (details explained in Section~\ref{sec:cases}). This yields $X_{FR}$ 
matrix of size 47$\times$80.

We decompose the matrix via RPCA s.t. $X_{FR}$ = $L_{FR}$ + $S_{FR}$. 
First we find that the rank of $L_{FR}$ is 26. This is exactly the number of 
unique geolocation and AS pairs of the server nodes in the rows of $X_{FR}$. To investigate this finding further, 
we randomly extract 500 submatrices of random sizes from $X_{FR}$. We decompose each submatrix 
via RPCA and compute the rank of their $L$ components. 
 
\begin{figure}[tbp]
\centerline{
\includegraphics[width=0.29\textwidth]{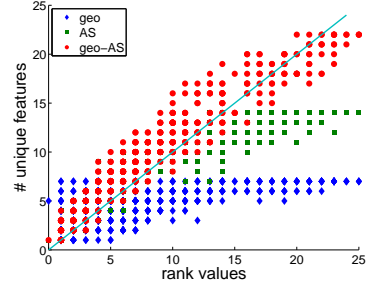}}
  \vspace{-0.04in}
\caption{The number of unique features of the server nodes and the destination prefixes vs. the rank values of the $L$ components of the randomly generated submatrices.} 
\label{fig:pfxlevel_rankL}
\vspace{-0.1in}
\end{figure}

Figure~\ref{fig:pfxlevel_rankL} plots the ranks of the $L$ components of the submatrices. We tag each row and column 
of the submatrices by their features, i.e. their geolocation, AS, and both. Then for each submatrix, we plot its rank vs. 
the minimum of the unique features on its rows and columns. Similar to our previous finding, the Figure~\ref{fig:pfxlevel_rankL}
shows that the number of unique geolocation and AS pairs in the rows/columns correlates well with the rank.

This analysis shows that the underlying dimensionality of the delay space is the result of routing choices as the paths that are destined to the
same prefixes in the same geolocation and the ASes tend to experience similar latencies. 



\section{Detecting Anomalies}  \label{sec:cases}  

Our goal is to identify anomalous routing paths via spectral analysis of latency values. 
Note that anomalous routing paths are the ones that have unexpectedly high RTTs 
due to routing misconfigurations or any suboptimal routing decisions
including the ones for load balancing.

\mpara{Aggregating RTT data.}
A given RTT value is composed of three components - transmission delay,
propagation delay, and queueing delay. Since each ICMP packet is 32 bytes, 
the transmission delays in our measurements are very small and can be ignored.
Therefore the delay we see is either propagation delay or queueing delay.

Since our goal is to find routing anomalies, we are interested in high propagation delays 
rather than high queuing delays. One way to eliminate the measurements with high queueing 
delays is to aggregate individual client IPs to their BGP prefixes as follows: For each server node 
we use the minimum RTT measured to any of the client IPs which belong to that prefix.  
We consider prefixes that have at least 10 IPs mapped to them so that
we significantly decrease the likelihood of queuing delays. Then, we generate $X_{FR}$ 
(also described in Section~\ref{sec:rank}), where each column represents a prefix and 
its entries are the minimum latencies from the server nodes 
to any of the client IPs in the prefix. We use $X_{FR}$ in anomaly detection
as presented below.



\mpara{Detecting Inflated Paths via Ratio Filtering.}
We decompose a latency matrix $X$ into its $L$ and $S$. As we explain 
in Section~\ref{sec:rpca}, the entries in $S$ represent the inflation that does
not fit into the low-rank structure of the latency measurements. In other 
words, the entries in $S$ are the difference between the measured latency in $X$ 
and the expected latency in $L$. 

Having decomposed the matrix, we say that the path from the server node $i$ to 
prefix $j$ is \emph{inflated} if the ratio of the inflation to the expected latency is 
greater than a threshold $\tau$, i.e. $\frac{S(i,j)}{L(i,j)} > \tau$.
In our analysis we set $\tau$ to 1. In other words, we investigate a route as an anomaly candidate 
if half of its measured RTT is estimated as inflation. Then we rank the candidates based on
the magnitude of $S(i,j)$ since larger the inflation, the most likely the path is
anomalous. In practice we find that paths whose inflation is more than 10 ms are worth investigating.


In addition, we find that, on the paths that are cross-continents, the inflation ratio is 
hardly greater than 1 as the minimum possible RTT is already high. For instance, in our dataset,
the minimum latency between Tokyo and Paris is 210 ms. and the minimum
latency between San Jose, US and Paris is 136.4 ms.
Therefore we also investigate all routes on which the magnitude of the 
inflation $S(i,j)$ is estimated greater than 30 ms.  



Next we present three cases of anomalies we detect in our dataset. In the first two cases, 
we apply our algorithm to the submatrices of $X_{FR}$ s.t. in each submatrix, columns are
prefixes from the same AS and rows are the server nodes which have measurements to all prefixes on the
columns. This guarantees that a) the rank of the underlying $L$ is very low (ranks vary 1-5) 
since the prefixes from the same AS and geolocation (in our case, France) are expected to have 
very similar columns, 
b) our results are not biased by the missing measurements.
We apply our filter to all such submatrices and below present two of them that have anomalies.
In the third case, we apply our method to entire $X_{FR}$ to test it 
against missing measurements and relatively higher rank value, i.e. the rank of $L_{FR}$ is 26. 
We present even in the case of missing measurements we can detect anomalies.

\begin{figure*}[tbp]
\centerline{
\includegraphics[width=0.30\textwidth]{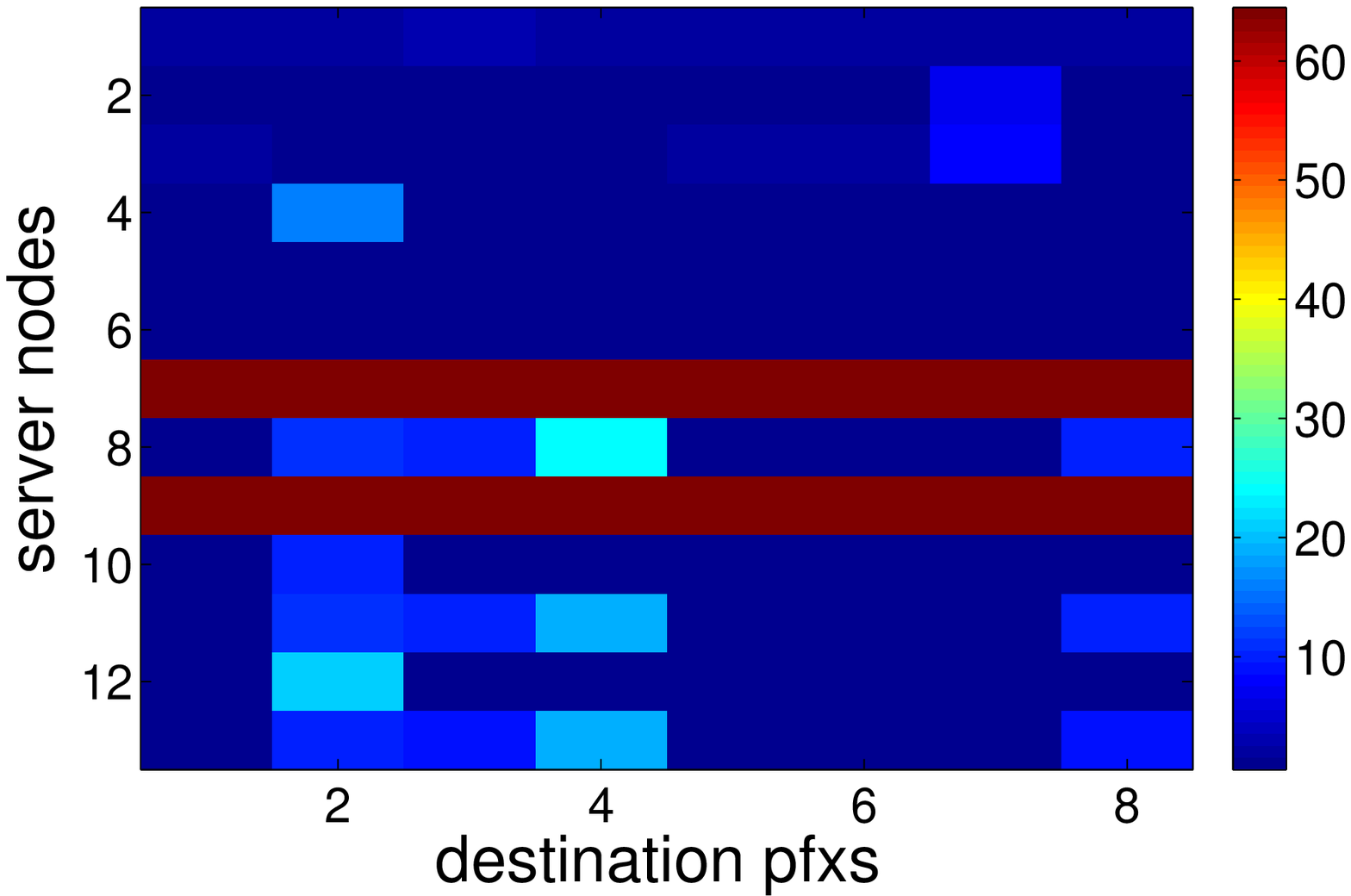}
\includegraphics[width=0.30\textwidth]{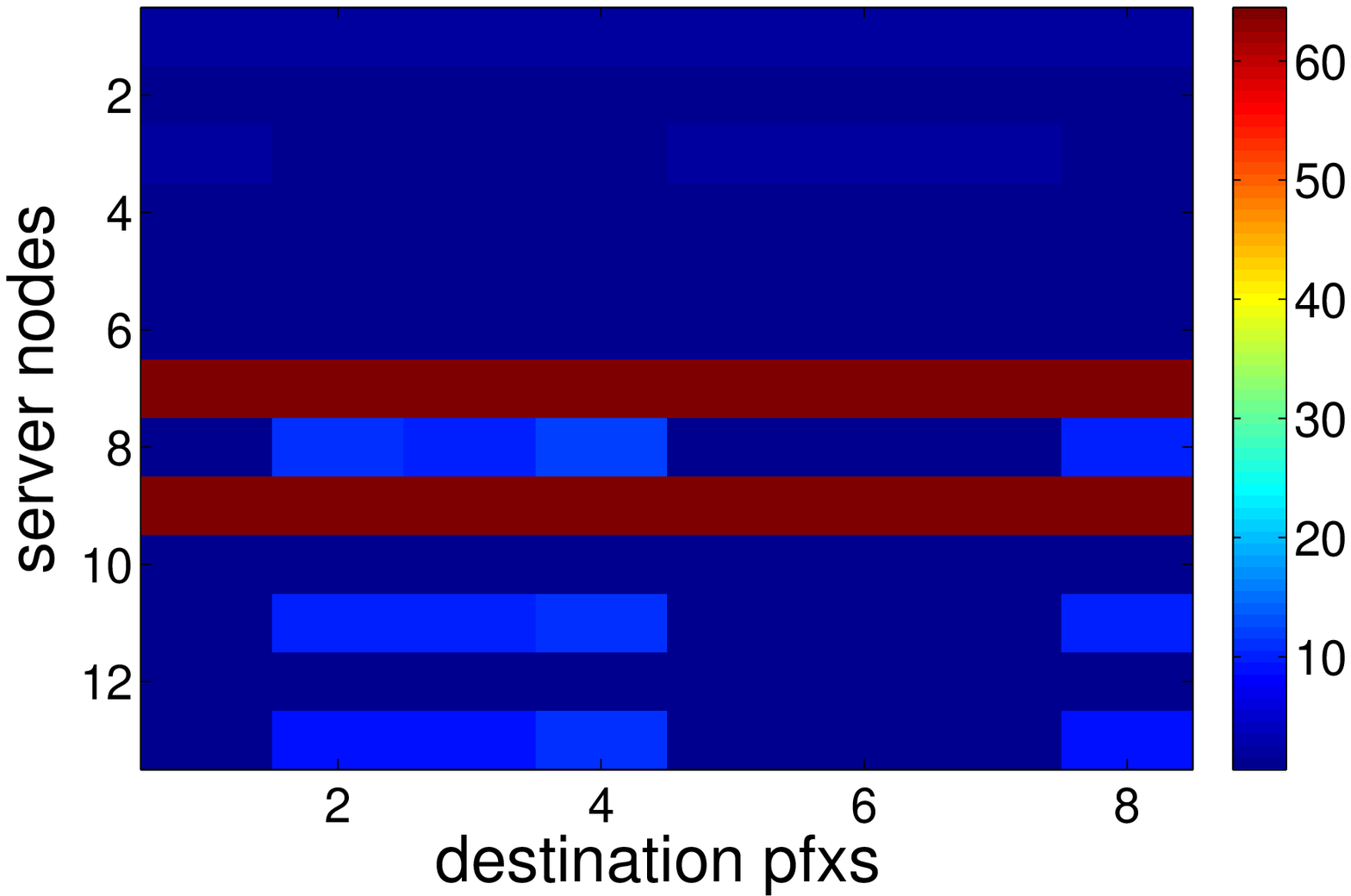}
\includegraphics[width=0.30\textwidth]{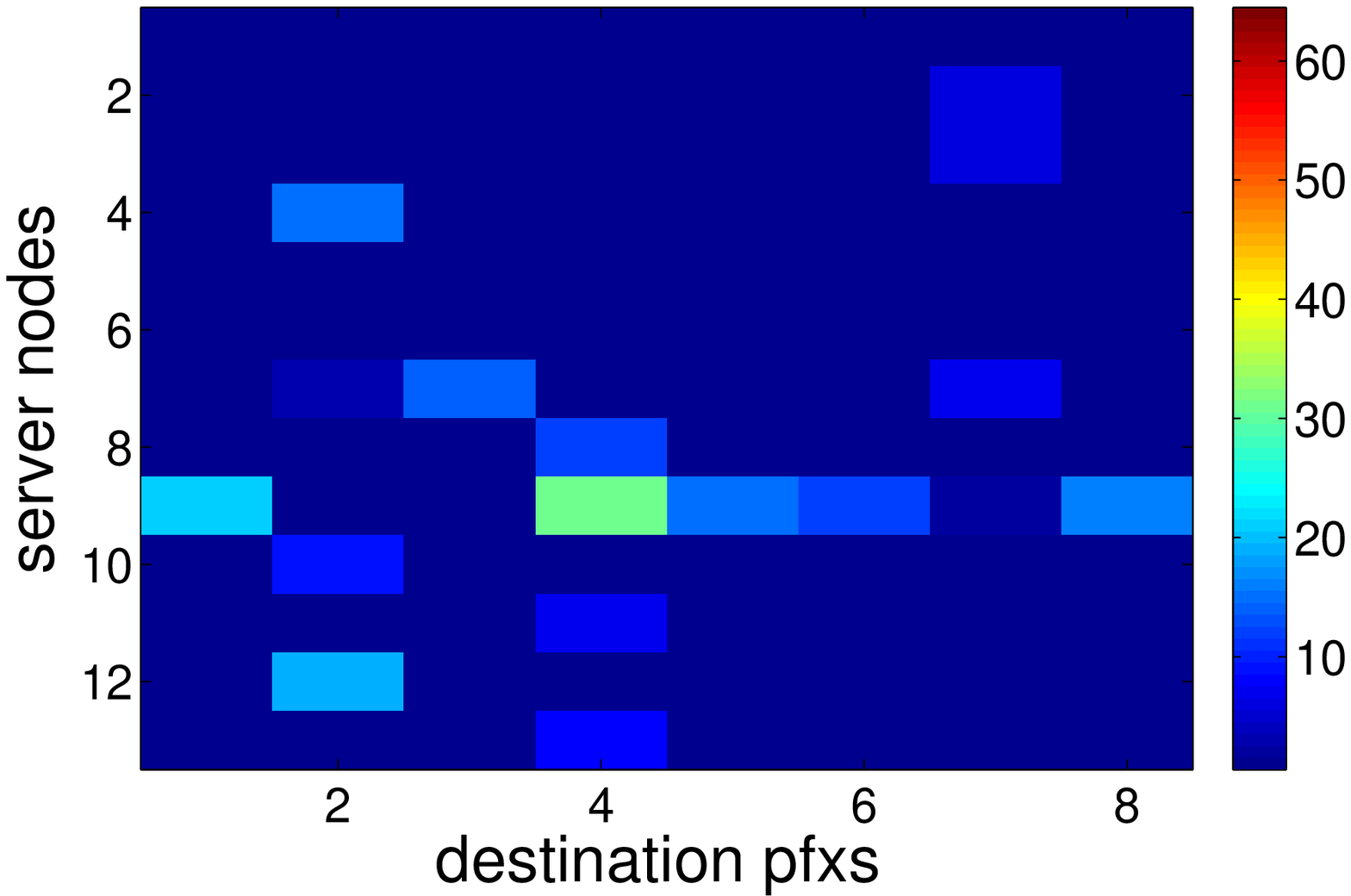}}
  \centerline{\hfill\hfill(a) $X$ \hfill\hfill\hfill(b) $L$ \hfill\hfill\hfill(c) $S$ \hfill\hfill}
    \vspace{-0.05in}
\caption{Latency values for Case 1 in $X$ (measured latency) and its decompositions, $L$ (expected latency) and $S$ (inflated latency).} 
\label{fig:FR_34164}
\end{figure*}


\mpara{Case 1. Detecting the anomalous path from a server node when the latency to only one destination prefix is inflated.}
In this case we study the RTT values to 8 prefixes that belong to Akamai International (AS34164). 
There are 13 server nodes in our dataset that have measurements to all of these prefixes, i.e. $X$ is 13$\times$8 as shown in 
Figure~\ref{fig:FR_34164}(a). The prefixes are listed in~Table\ref{tab:FR_34164} by the order they appear in the columns of Figure~\ref{fig:FR_34164}. 
The heatmaps in Figure~\ref{fig:FR_34164} visualize the measured latency values in $X$ as well as the latencies in the $L$ and $S$ components. 
Below are the four anomalies we find in this case.

\begin{table}
   \centering
   \resizebox{0.8\columnwidth}{!}{
   \begin{tabular}{|c|l|c|l|}
    \hline
        Column id & Prefix & Column id & Prefix  \\ \hline
        1   & 184.85.251.0/24  &      5   & 2.18.249.0/24  \\ 
        2   & 2.16.126.0/24    &   6   & 23.62.9.0/24   \\ 
        3   & 2.16.136.0/24      &  7   & 92.123.193.0/24   \\
   	4   & 2.16.54.0/24      &   8   & 96.16.122.0/23   \\ 
    \hline   	
    \end{tabular}}
    \caption{Prefix list for Case 1. Prefixes belong to AS34164. The prefixes appear in the order of column id in Figure~\ref{fig:FR_34164}.}
    \label{tab:FR_34164}
    \vspace{-0.2in}
\end{table}

The first anomalous route is from a server node in Deutsche Telekom (AS3320) in Paris to 2.16.126.0$/$24.
Their path corresponds to row 4 and column 2 of the matrices in Figure~\ref{fig:FR_34164}. The RTT on the path is
16.2 ms and our analysis estimates that the RTT should have been 0.9 ms (see Figure~\ref{fig:FR_34164}(b))
and therefore it is inflated by 15.3 ms (see Figure~\ref{fig:FR_34164}(c)). 

Looking at the traceroute path from the server node to the prefix, we see that the inflation is due to
a detour through Munich and Milano, i.e. the route to the prefix is AS3320 (Paris) $\rightarrow$ AS3320 (Munich) 
$\rightarrow$ AS6762 (Milano) $\rightarrow$ AS6762 (Paris) $\rightarrow$ AS34164 (Aubervilliers)
\footnote{We map all routers on a given traceroute path to their AS and geolocation. Then, we represent the consecutive 
routers that have the same AS-geolocation tag as one hop for readability.}. 
Looking at the traceroutes of the other seven prefixes, we do not see such detour. Their routes are 
AS3320 (Paris) $\rightarrow$ AS3257 (Paris) $\rightarrow$ AS34164 
(Aubervilliers or Paris). 
This indicates that there might be a misconfiguration for the prefix 2.16.126.0$/$24. 
In addition, looking at our BGP data, we see that AS3320 and AS34164 are peers for many other locations
in Europe (Australia, Belgium, Italy, Russia, Netherlands etc.) and the direct link between them is used 
for the prefixes in these locations. Therefore we infer that shorter routes might be possible for the prefix 2.16.126.0$/$24
if the peering relationship between AS3320 and AS34164 is used. 

The second anomalous route is to the same prefix, \\
2.16.126.0$/$24 from a server node in Nerim SAS (AS13193) in Paris.
Their path corresponds to row 12 and column 2 of the matrices in Figure~\ref{fig:FR_34164}. The RTT on the path is
21.1 ms and our analysis estimates that the RTT should have been 1.1 ms (see Figure~\ref{fig:FR_34164}(b))
but it is off by 20 ms (see Figure~\ref{fig:FR_34164}(c)). 

Looking at the traceroute path from the server node to the prefix, we see that the reason for 20 ms inflation is
because a detour through Dublin, London, and Amsterdam, i.e. the path is 
AS13193 (Paris) $\rightarrow$ AS10310 (Dublin) $\rightarrow$ AS10310 (London) $\rightarrow$ AS5580 (London) 
$\rightarrow$ AS5580 (Amsterdam) $\rightarrow$ AS34164 (Aubervilliers). 
Looking at the traceroutes of the other seven prefixes, we do not see such detour. Their paths from the same server node
go direct, i.e. AS13193 (Paris) $\rightarrow$ AS1299 (Paris) $\rightarrow$ AS5580 (Aubervilliers) $\rightarrow$
AS34164 (Aubervilliers). Therefore we infer that there is a misconfiguration for prefix 2.16.126.0$/$24.

The third anomalous route is from a server node in Orange Telecom (AS 5511) in Paris to 2.16.54.0$/$24. 
Their path corresponds to row 8 and column 4 of the matrices in Figure~\ref{fig:FR_34164}. The RTT on the path is
24.9 ms and our analysis estimates that the RTT should have been 12.3 ms (see Figure~\ref{fig:FR_34164}(b))
but it is off by 12.6 ms (see Figure~\ref{fig:FR_34164}(c)). 

Looking at the traceroute path from the server node to the prefix, we see that the inflation is due to
a detour via Frankfurt, i.e. the path is 
AS5511 (Paris) $\rightarrow$ AS3257 (Frankfurt) $\rightarrow$ AS3257 (Paris) $\rightarrow$ AS34164 (Aubervilliers).
However, looking at the paths from the same server node to the other prefixes, we see that there is a direct path
AS5511 (Paris) $\rightarrow$ AS5511 (Aubervilliers) $\rightarrow$ AS34164 (Aubervilliers). 


The fourth anomalous route is from a server node in NTT Comm. (AS2914) in Tokyo to 2.16.54.0$/$24. 
This path corresponds to row 9 and column 4 of the matrices in Figure~\ref{fig:FR_34164}. We identify that the 
path is inflated because of a detour Hong Kong, Singapore, and Mumbai. We identify a very similar case and discuss it in detail in the next case. 



\begin{figure*}[tbp]
\centerline{
\includegraphics[width=0.30\textwidth]{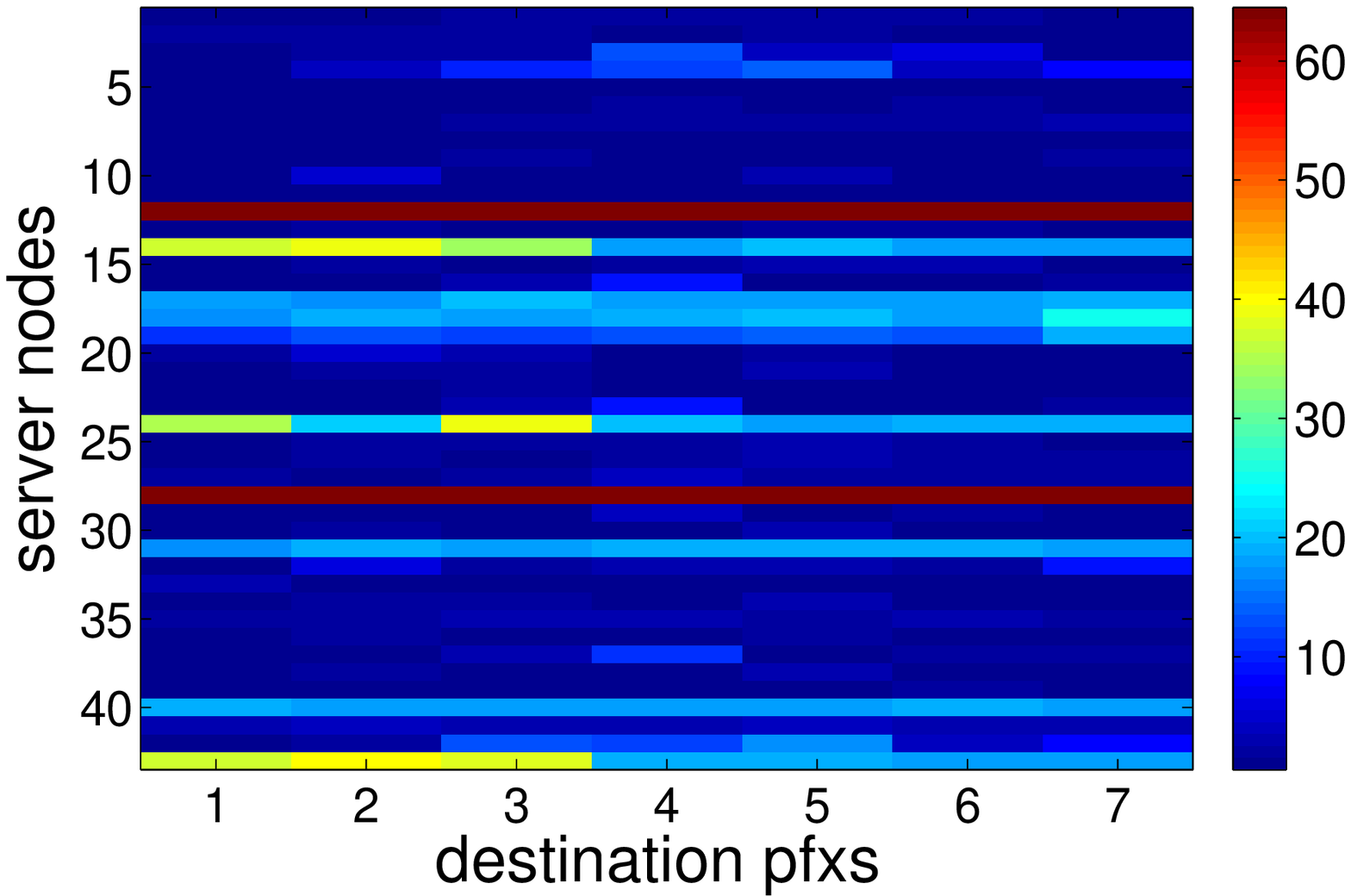}
\includegraphics[width=0.30\textwidth]{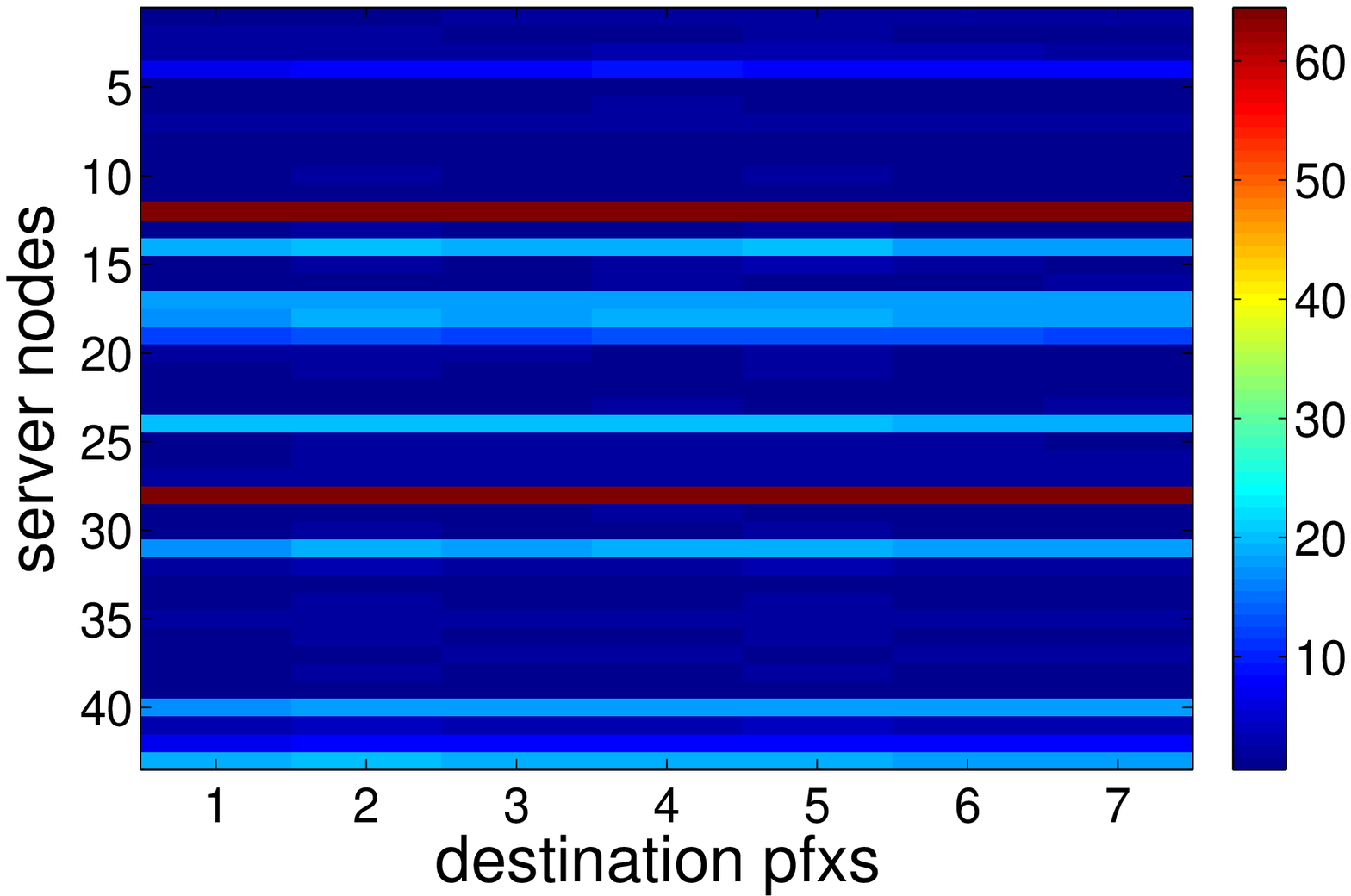}
\includegraphics[width=0.30\textwidth]{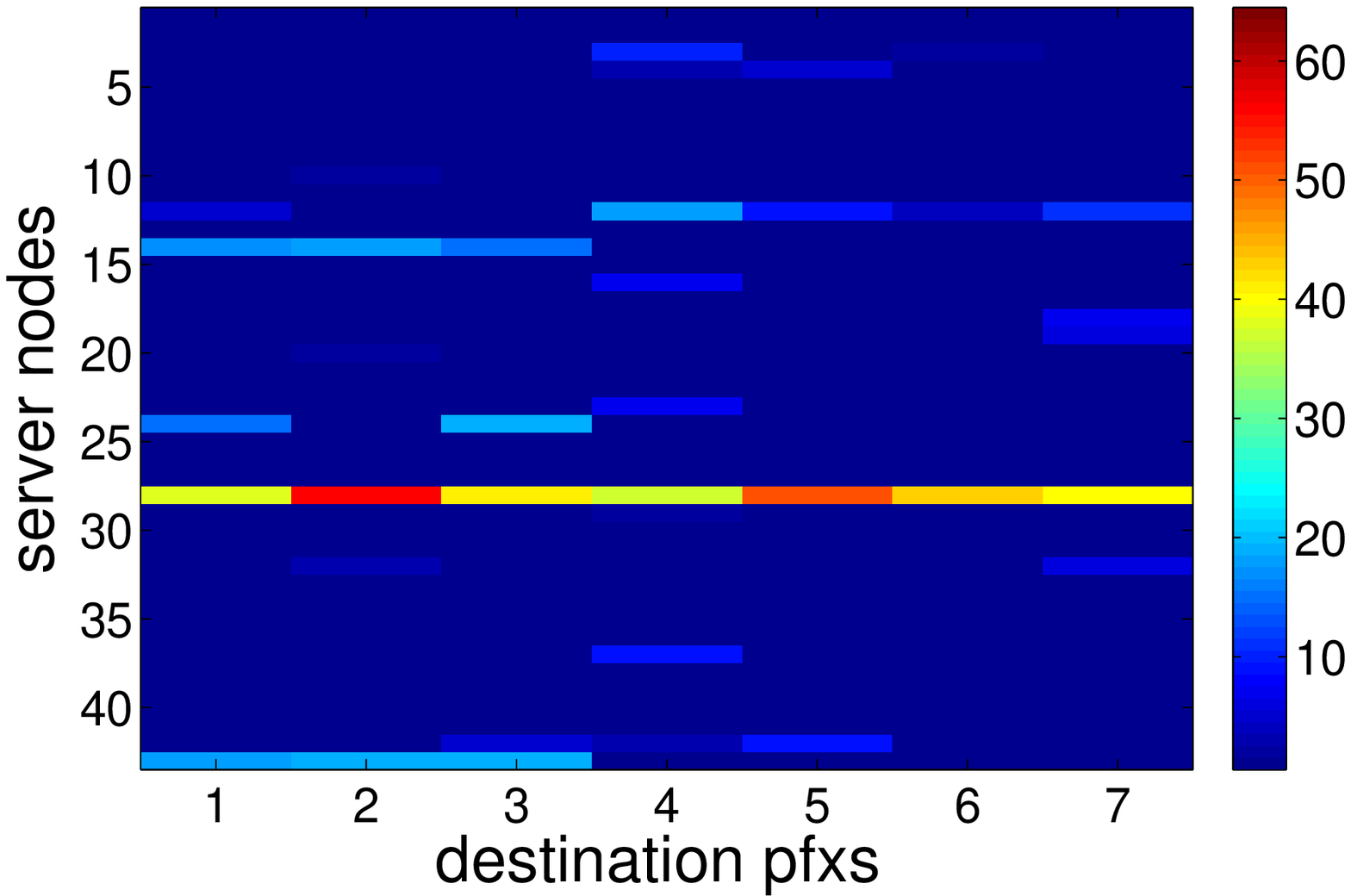}}
  \centerline{\hfill\hfill(a) $X$ \hfill\hfill\hfill(b) $L$ \hfill\hfill\hfill(c) $S$ \hfill\hfill}
    \vspace{-0.05in}
\caption{Latency values for Case 2 in $X$ (measured latency) and its decompositions, $L$ (expected latency) and $S$ (inflated latency).} 
\label{fig:FR_12670}
\vspace{-0.1in}
\end{figure*}

\begin{table}
   \centering
      \resizebox{0.8\columnwidth}{!}{
   \begin{tabular}{|c|l|c|l|}
    \hline
        Column id & Prefix &  Column id & Prefix \\ \hline
        1   & 195.167.192.0/20     & 5   & 89.225.192.0/18 \\ 
        2   & 212.99.0.0/17      & 6   & 92.103.0.0/16\\ 
        3   & 46.218.0.0/16      & 7   & 92.103.64.0/18\\
   	4   & 46.218.0.0/18      &     &   \\
    \hline   	
    \end{tabular}}
    \caption{Prefix list for Case 2. Prefixes belong to AS12670. The prefixes appear in the order of column id in Figure~\ref{fig:FR_12670}.}
    \label{tab:FR_12670}
     \vspace{-0.2in}
\end{table}

\mpara{Case 2. Detecting anomalies even when the paths to all prefixes from the same AS and geolocation are inflated.}
This case presents that our method can catch anomalies even when all the prefixes within the same AS and
geolocation are infected. In this case we study the RTT values to 7 prefixes that belong to CompleTel (AS12670) located in France. 
There are 13 server nodes in our dataset that have measurements to all of these prefixes, i.e. $X$ is 43$\times$7 as shown in 
Figure~\ref{fig:FR_12670}(a). The prefixes are listed in Table~\ref{tab:FR_12670} by the order they appear in the columns of Figure~\ref{fig:FR_12670}. 

The first anomalous route is between a server node from NTT Communications (AS2914) in Tokyo to all prefixes
in the list. This server node corresponds to row 28 in Figure~\ref{fig:FR_12670}. The RTT values from this server
node to the prefixes vary 250ms - 280ms. Figure~\ref{fig:FR_12670}(b) estimates RTT values should be in the 
210-222 ms range, therefore they are all inflated by 40-56 ms as shown in row 28 of Figure~\ref{fig:FR_12670}(c).

Looking at the traceroute paths from the server node to the prefixes, we find that all the paths are longer than expected.
The paths are  AS2914 (Tokyo) $\rightarrow$ AS2914 (HongKong) $\rightarrow$  AS6453 (HongKong)
$\rightarrow$ AS6453 (Singapour) $\rightarrow$ AS6453 (Mumbai)
$\rightarrow$ AS6453(Marseille) $\rightarrow$ AS6453(Paris)
$\rightarrow$ AS12670 (Paris). 

In order to find whether there is a shorter path between AS2914 (Tokyo) and  AS12670 (Paris), 
we investigate all paths in between them. We find that AS2914 has presence in many locations in 
Europe including Paris and therefore shorter paths between AS2914 (Tokyo) and  AS12670 (Paris, 
without redirection via AS6453, are possible. Alternatively, we find that AS2914 makes different next hop decisions
for some other prefixes that are also located in Paris. For instance, 
for the prefix 83.167.32.0$/$19 of AS8218 (Neo Telecoms), there is the following path,   
AS2914 (Tokyo) $\rightarrow$ AS2914 (Seattle) $\rightarrow$ AS6461 (Seattle) $\rightarrow$ AS6461 (Chicago) $\rightarrow$ 
AS6461 (New York) $\rightarrow$ AS6461 (Paris) $\rightarrow$ AS8218 (Paris).
This path is of 210 ms as opposed to the ones via AS6453 that are 250-280 ms. 
In our BGP data, we find that AS6461 has a peering with AS12670 and could be chosen 
by AS2914 as an alternative to AS6453. This would lead to a smaller RTT route. We note that 
the decision of routing via AS6453 might be due to load balancing. 


The second anomalous path is from a server node in Orange Telecom (AS5511) in Aubervilliers, France
to the first three prefixes listed in Table~\ref{tab:FR_12670}. This server node corresponds to row 43 in Figure~\ref{fig:FR_12670}.
The RTT values from this server node to the prefixes vary 37.7 - 40.3 ms. Figure~\ref{fig:FR_12670}(b) estimates RTT values should be in the 19.3-20.4 ms range, therefore they are all inflated by around 19ms as shown in row 28 of Figure~\ref{fig:FR_12670}(c).

The traceroute paths from the server node to these three prefixes are 
AS5511 (Aubervilliers) $\rightarrow$ AS5511 (Paris) $\rightarrow$ AS174 (Paris) $\rightarrow$ AS12670 (Paris).
However, the paths to the other four prefixes  has lower RTT values, i.e. 
AS5511 (Aubervilliers) $\rightarrow$ AS5511 (Paris) $\rightarrow$ AS6453 (Paris) $\rightarrow$ AS12670 (Paris). 
Although both paths are 4 hops, the latency on the link between AS5511 and AS6453 is less than the latency 
on the link between AS5511 and AS6453. Notice that 46.218.0.0$/$18 is the specific subset of 46.218.0.0$/$16 
and follow a shorter path.


\begin{figure*}[tbp]
\centerline{
\includegraphics[width=0.31\textwidth]{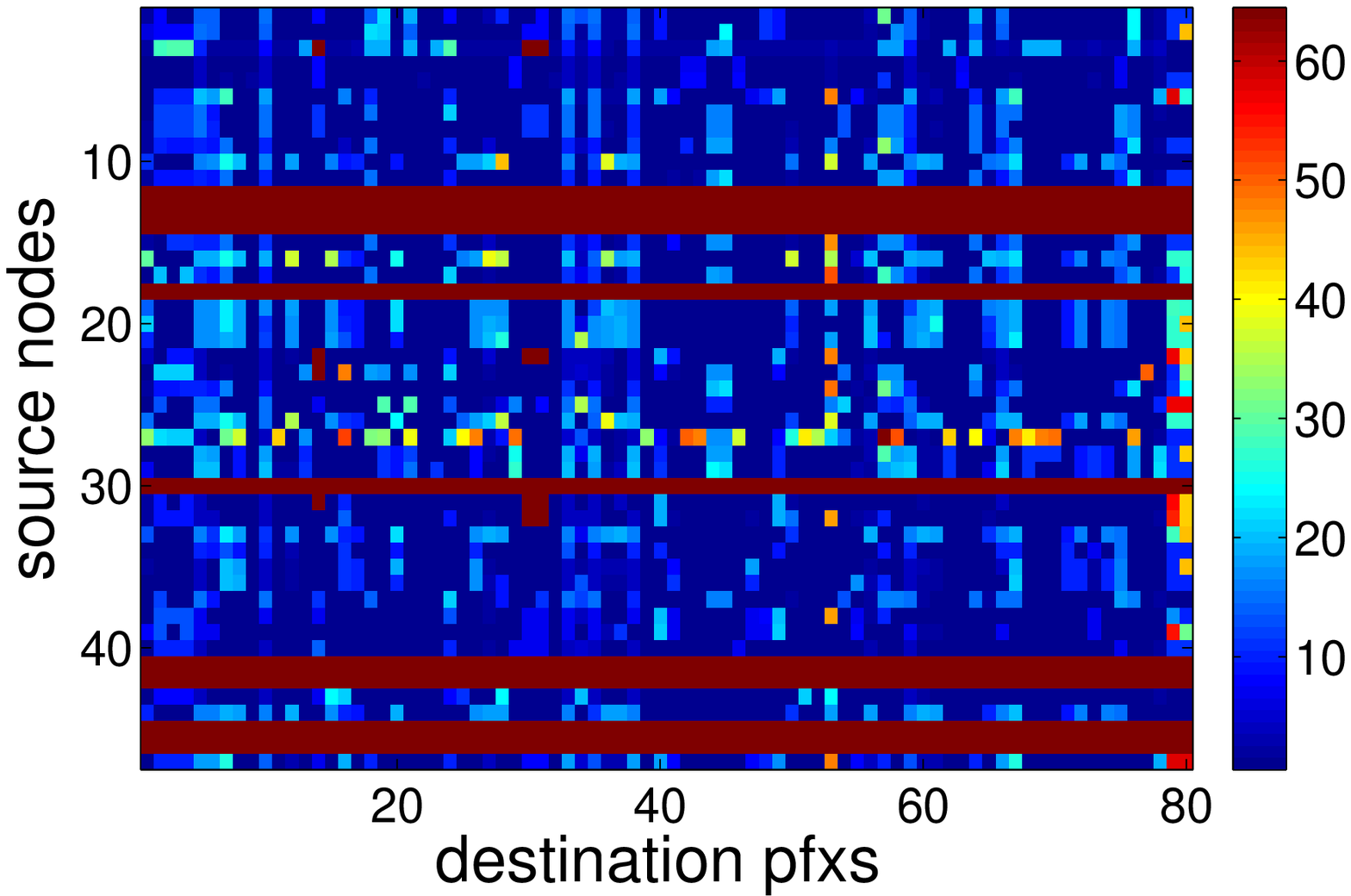}
\includegraphics[width=0.31\textwidth]{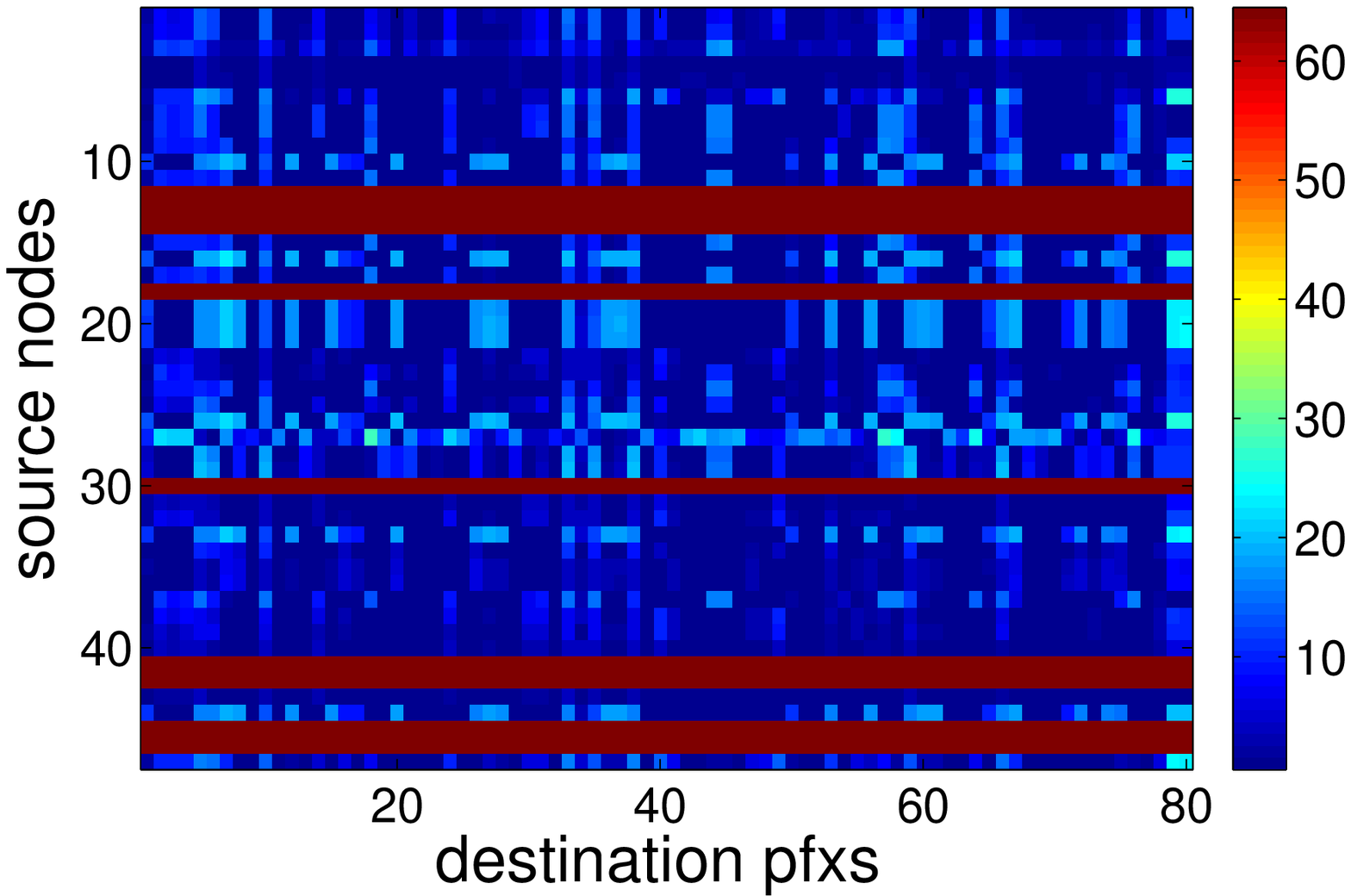}
\includegraphics[width=0.31\textwidth]{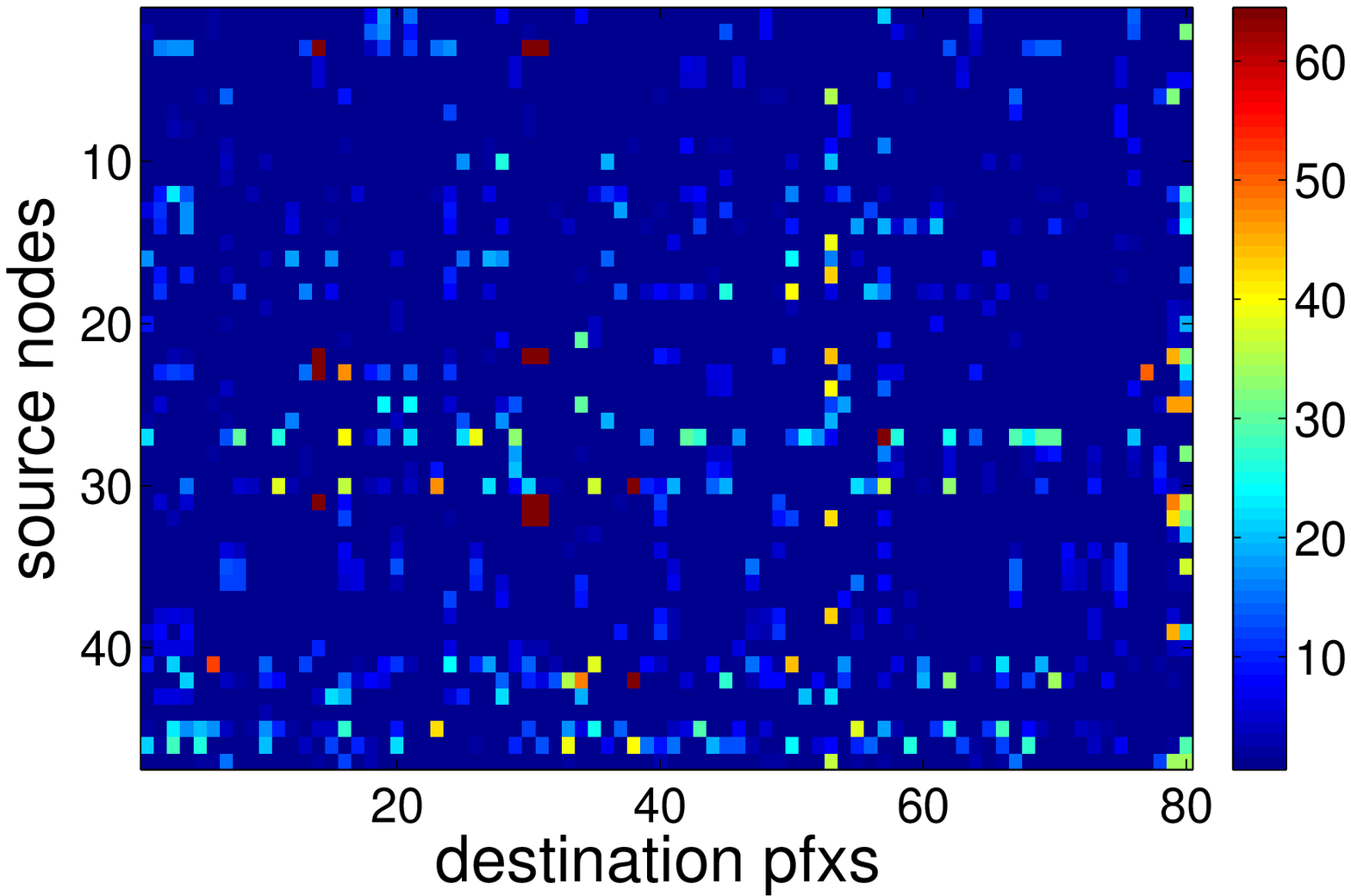}}
  \centerline{\hfill\hfill(a) $X_{FR}$ \hfill\hfill\hfill(b) $L_{FR}$ \hfill\hfill\hfill(c) $S_{FR}$ \hfill\hfill}
  \vspace{-0.09in}
\caption{Latency values for Case 3 in $X_{FR}$ (measured latency) and its decompositions, $L_{FR}$ (expected latency) and $S_{FR}$ (inflation).} 
\label{fig:FR_full}
\vspace{-0.19in}
\end{figure*}

\mpara{Case 3. Detecting anomalies in the case of missing measurements.}
First two cases test our method when the RTTs between all end hosts are known.
Finally, we show how to apply our method in the presence of missing values. 

As a preprocessing step, we interpolate the missing measurements as follows.
For each missing measurement from a server node $i$ to a prefix $j$, we replace the missing measurement
with the minimum RTT value observed from another server node that is in the same location and
AS as the server node $i$ to any prefix that is in the same AS and location as the prefix $j$. Such 
interpolation replaces the 0-valued missing entries with the lowest RTT seen between two
AS-geolocation regions. This preprocessing step decreases the ratio of missing measurements in 
$X_{FR}$ from 15\% to 1\%. 

Next we apply RPCA  on the interpolated matrix, followed by the same inflation filter. 
Figure~\ref{fig:FR_full} shows the latency values. 
We find that the server node that has the most anomalous routes is located in Akamai (AS20940) in Aubervilliers.
This server node corresponds to row 27 and it has 27 anomaly candidates as shown in Figure~\ref{fig:FR_full}(b-c).
We find that the paths from this server node make detour via Amsterdam and Zurich although the both the server node and the prefixes are in France.  
For instance, to prefix 2.20.243.0$/$24 (column 53), the measured RTT is 21.94 ms. However, the
estimated latency in $L_{FR}$ is 9.87 ms and the estimated inflation in $S_{FR}$ is 12.07 ms. 
The traceroute path to the prefix is 
AS20940 (Aubervilliers) $\rightarrow$ AS12322 (Paris) $\rightarrow$ AS1200 (Amsterdam) 
$\rightarrow$ AS13030 (Amsterdam) $\rightarrow$ AS13030 (Zurich) $\rightarrow$ AS20940 (Aubervilliers). 

This case shows that our method catches anomalous paths even when the magnitude of RTT is relatively low - that is when the
inflation is not obvious. The average latency from this server node to all prefixes is 18.3 ms. 
That is, comparing the RTT value of the anomalous path, 21.94 ms,  with the rest of the paths 
from the same server node to the same region would not detect the anomaly. However, our method successfully pinpoints the 
anomalous paths by the inflation-to-estimated-latency ratio filter. 

In conclusion, we show that with the help of a simple interpolation step, our method
successfully detects anomalies even when a significant portion of the RTTs are missing.
\vspace{-0.1in}



\section{Related Work}  \label{sec:rel-work}    
Understanding the delay space and detecting RTT inflations are of great interest in the literature. 
\cite{Tangmunarunkit01, Savage99} show that half of the paths are inflated due to routing policies.
\cite{Spring03} studies the possible root causes of path inflation and show that intradomain 
routing decisions and peering policies are the major reasons. \cite{Pucha07} studies the impact of routing changes 
on network delay and jitter using network topology. Similar to our findings \cite{Pucha07} shows that intradomain 
routing decisions can cause as severe latency inflation as interdomain routing decisions.
\cite{Muhlbauer10} studies how routing parameters impact path optimality.
\cite{Zarifis14} studies the causes of path inflation in mobile networks.
Our work is complementary to all these studies as our method provides 
a list of anomaly candidate paths to investigate.

The low-rank property of latency matrices is used in various studies.
\cite{Dabek04, Madhyastha06} propose methods for predicting RTT based on this observation. 
\cite{Tang03} uses dimensionality reduction as a way of estimating RTTs between 
hosts without direct measurements. \cite{Liao12, Liao13} uses the low-rank property as a way to infer 
proximity between hosts. \cite{Mao04} presents a latency estimation system that uses matrix factorization 
which also uses the low-rank structure of latency data. In addition, \cite{Zhang06, Abrahao08, Wong05, Lim03} show 
that latency space can be embedded into low-dimensional coordinate spaces. All these work are similar to 
ours as they leverage the highly structured nature of the latency data. Unlike these work, we use the low-rank
property to detect the routing anomalies in the Internet.
\vspace{-0.1in}


\section{Conclusion}   \label{sec:conc}   In this paper, we study the delay space of the Internet via robust principal component analysis. 
We find that the dimensionality of the delay space 
is well-correlated with the geolocation and ASes of the end hosts.
Then we show how to leverage low-rank property of the delay space to 
identify anomalous routing paths. We show that our 
method successfully identifies the anomalies even when all prefixes
from the same region are infected and in the presence of missing
latency measurements. 


\bibliographystyle{plain}
\bibliography{paper}

\end{document}